\documentclass[twocolumn,floatfix]{aastex62}

\usepackage{natbib}
\usepackage{amsmath}
\setcitestyle{sort,comma}
\usepackage[colorinlistoftodos]{todonotes}
\usepackage{textcase}
\usepackage{hyperref}

\graphicspath{{./}{figures/}}

\submitjournal{ApJ}

\shorttitle{New Insights into the Keystone WASP-107 System}

\shortauthors{Piaulet et al.}

\begin{document}

\title{WASP-107b's density is even lower: a case study for the physics of planetary gas envelope accretion and orbital migration}









\correspondingauthor{Caroline Piaulet}
\email{caroline@astro.umontreal.ca}

\author[0000-0002-2875-917X]{Caroline Piaulet}
\affil{Department of Physics, and Institute for Research on Exoplanets, Universit\'{e} de Montr\'{e}al, Montreal, H3T 1J4, Canada}

\author[0000-0001-5578-1498]{Bj\"{o}rn Benneke} 
\affil{Department of Physics, and Institute for Research on Exoplanets, Universit\'{e} de Montr\'{e}al, Montreal, H3T 1J4, Canada}

\author[0000-0003-3856-3143]{Ryan A. Rubenzahl}
\affil{Cahill Center for Astrophysics, California Institute of Technology, 1216 East California Boulevard, Pasadena, CA 91125, USA}

\author[0000-0001-8638-0320]{Andrew W. Howard}
\affil{Cahill Center for Astrophysics, California Institute of Technology, 1216 East California Boulevard, Pasadena, CA 91125, USA}

\author{Eve J. Lee}
\affil{Department of Physics and McGill Space Institute, McGill University, 3550 rue University, Montreal, QC, H3A 2T8, Canada}

\author[0000-0002-5113-8558]{Daniel Thorngren}
\affil{Department of Physics, and Institute for Research on Exoplanets, Universit\'{e} de Montr\'{e}al, Montreal, H3T 1J4, Canada}

\author[0000-0003-4540-5661]{Ruth Angus}
\affil{Department of Astrophysics, American Museum of Natural History, 200 Central Park West, Manhattan, NY, USA}
\affil{Center for Computational Astrophysics, Flatiron Institute, 162 5th Avenue, New York, NY, 10010, USA}

\author{Merrin Peterson}
\affil{Department of Physics, and Institute for Research on Exoplanets, Universit\'{e} de Montr\'{e}al, Montreal, H3T 1J4, Canada}

\author{Joshua E. Schlieder}
\affil{NASA Goddard Space Flight Center, 8800 Greenbelt Road, MD, USA}

\author{Michael Werner}
\affil{Jet Propulsion Laboratory, California Institute of Technology, 4800 Oak Grove Drive, Pasadena, CA 91109, USA}

\author[0000-0003-0514-1147]{Laura Kreidberg}
\affiliation{Max Planck Institute for Astronomy, K{\"o}nigstuhl 17, 69117 Heidelberg, Germany}

\author{Tareq Jaouni}
\affil{Department of Physics, University of Ottawa, 25 Templeton St., Ottawa, ON, K1N 6N5, Canada}

\author{Ian J.~M.\ Crossfield}
\affiliation{Department of Physics \& Astronomy, University of Kansas, 1082 Malott, 1251 Wescoe Hall Dr., Lawrence, KS 66045, USA}

\author[0000-0002-5741-3047]{David R. Ciardi}
\affil{Caltech/IPAC-NASA Exoplanet Science Institute, 770 S. Wilson Ave, Pasadena, CA 91106, USA}

\author{Erik A. Petigura}
\affil{Department of Physics \& Astronomy, University of California Los Angeles, Los Angeles, CA 90095, USA}

\author[0000-0002-4881-3620]{John Livingston}
\affil{Department of Astronomy, University of Tokyo, 7-3-1 Hongo, Bunkyo-ku, Tokyo 113-0033, Japan}

\author{Courtney D.\ Dressing}
\affil{Department of Astronomy, University of California, Berkeley, CA 94720}

\author{Benjamin J. Fulton}
\affil{NASA Exoplanet Science Institute, California Institute of Technology, 1200 East California Blvd, Pasadena, CA 91125, USA}

\author{Charles Beichman}
\affil{Jet Propulsion Laboratory, California Institute of Technology, 4800 Oak Grove Drive, Pasadena, CA 91109, USA}
\affil{Caltech/IPAC-NASA Exoplanet Science Institute, 770 S. Wilson Ave, Pasadena, CA 91106, USA}

\author{Jessie L.\ Christiansen}
\affil{Caltech/IPAC-NASA Exoplanet Science Institute, 770 S. Wilson Ave, Pasadena, CA 91106, USA}

\author{Varoujan Gorjian}
\affil{Jet Propulsion Laboratory, California Institute of Technology, 4800 Oak Grove Drive, Pasadena, CA 91109, USA}

\author[0000-0003-3702-0382]{Kevin K.\ Hardegree-Ullman}
\affil{Caltech/IPAC-NASA Exoplanet Science Institute, 770 S. Wilson Ave, Pasadena, CA 91106, USA}

\author{Jessica Krick}
\affil{Caltech/IPAC-NASA Exoplanet Science Institute, 770 S. Wilson Ave, Pasadena, CA 91106, USA}

\author{Evan Sinukoff}
\affil{Cahill Center for Astrophysics, California Institute of Technology, 1216 East California Boulevard, Pasadena, CA 91125, USA}
\affil{Institute for Astronomy, University of Hawai‘i at Mānoa, Honolulu, HI 96822, USA}



\begin{abstract}

With a mass in the Neptune regime and a radius of Jupiter, WASP-107b presents a challenge to planet formation theories. Meanwhile, the planet's low surface gravity and the star's brightness also make it one of the most favorable targets for atmospheric characterization.
Here, we present the results of an extensive 4-year Keck/HIRES radial-velocity (RV) follow-up program of the WASP-107 system and provide a detailed study of the physics governing the accretion of its gas envelope. We reveal that WASP-107b's mass is only 1.8 Neptune masses ($M_b = 30.5 \pm 1.7$~$M_\oplus$).
The resulting extraordinarily low density suggests that WASP-107b has a H/He envelope mass fraction of $> 85$\%  unless it is substantially inflated.
The corresponding core mass of $<4.6$ $M_\oplus$ at 3$\sigma$ is significantly lower than what is traditionally assumed to be necessary to trigger massive gas envelope accretion.
We demonstrate that this large gas-to-core mass ratio most plausibly results from the onset of accretion at $\gtrsim 1$ AU onto a low-opacity, dust-free atmosphere and subsequent migration to the present-day $a_b = 0.0566 \pm 0.0017$ AU.
Beyond WASP-107b, we also detect a second more massive planet ($M_c \sin i = 0.36 \pm 0.04 M_{J}$) on a wide eccentric orbit ($e_c = 0.28 \pm 0.07$) which may have influenced the orbital migration and spin-orbit misalignment of WASP-107b. Overall, our new RV observations and envelope accretion modeling provide crucial insights into the intriguing nature of WASP-107b and the system's formation history. Looking ahead, WASP-107b will be a keystone planet to understand the physics of gas envelope accretion.
\end{abstract}



\section{Introduction} \label{sec:intro}

Plentiful discoveries in exoplanetary science have revealed a great diversity of planetary systems, raising so far more new questions regarding the formation of planets than they have provided answers. The formation, evolution and orbital migration mechanisms of hot Jupiters and sub-Neptunes, for instance, which have no analogue around the sun, are challenging current theories that were originally developed solely based on Solar System observations \citep{Lee2016,Dawson2018}. Similarly, detailed investigations of exoplanets straddling the boundary between ice and gas giants, so-called ``super-Neptunes'' with masses intermediate between that of Neptune and Saturn, have the potential to reshape our understanding of planetary formation and migration (e.g. \citealp{Lee2019}).

WASP-107b is a super-puffy super-Neptune discovered by WASP-South \citep{Anderson2017}. Transit lightcurves obtained with WASP-South and \textit{K2}, combined with CORALIE RV measurements \citep{Anderson2017,DaiWinn2017} place this planet directly at the transition between the ice and gas giants of the solar system. Its mass was estimated to be $2.2\pm 0.2 M_{\mathrm{Nep}}$ or $38 \pm 3 M_\oplus$, but its radius is closest to that of Jupiter (0.94 R$_{J}$). This first mass measurement identified WASP-107b as having one of the lowest bulk densities among all the discovered extrasolar planets. This suggests that WASP-107b is a gaseous planet with most of its mass contained within its massive envelope consisting primarily of hydrogen and helium (the most abundant gases in protoplanetary disks). A more precise and accurate measurement of the mass of this planet would provide critical physical constraints both on the core mass necessary to accrete such a large atmosphere and on the gas accretion processes at play, while being essential for accurate atmosphere characterization \citep{Batalha2019}. 

WASP-107b is a challenge for planet formation as it harbors a large envelope yet currently orbits at a mere 0.06 AU from its star. 
WASP-107b's core accreted over 10 $M_\oplus$ in gas, suggesting that it formed at several AU from the host star where the protoplanetary disk is rich in gas, ices and dust particles that can be accreted onto the rocky core. Therefore, WASP-107b has likely undergone inwards migration. Whether this migration was driven by interactions with the protoplanetary disk (disk-driven migration, e.g. \citealp{Lin96}) or with other objects in the system (e.g. high-eccentricity pathway; \citealp{Nagasawa2008}) still remains unknown. Moreover, WASP-107b's low mass suggests that it did not undergo the full extent of runaway gas accretion. However, the core-nucleated accretion theory \citep{Pollack1996} predicts that planets that reach a core mass of about 10 $M_\oplus$ will undergo runaway gas accretion and reach a Jupiter mass within the disk's lifetime. This leaves us with two unanswered questions: \textit{How and where did WASP-107b form?} and \textit{Which process prevented it from reaching a Jupiter mass?}

The large atmospheric scale height and favorable radius of this extremely low density planet make it a very exciting target for atmospheric characterization through transmission spectroscopy. It was observed with the \textit{Hubble Space Telescope} (\textit{HST}) in transit in the \textit{WFC3}/G102 and G141 bandpasses, and the measured wavelength-dependent transit depth yielded a transmission spectrum in the $\approx 0.9-1.6~\mu$m region \citep{Kreidberg2018,Spake2018}. Spectral retrieval methods using these datasets have already unveiled peculiar features of the atmosphere of this planet. The presence of high-altitude aerosols was inferred from the muted spectroscopic signature of water (consistent with a solar abundance pattern; \citealt{Kreidberg2018}), and helium was detected, making WASP-107b the first exoplanet with a direct helium detection \citep{Spake2018,Allart2019}. However, all atmospheric studies so far relied on an uncertain and biased mass measurement (see Section \ref{ssec:mass_orb_res}), with direct implications on the assumed surface gravity and retrieval outcomes. Thus, a better mass determination is critical for interpreting spectroscopic observations.

Here, we present results from an extensive follow-up campaign of WASP-107 and detailed modeling of the accretion of WASP-107b's envelope. We describe the Keck/HIRES observations in Section \ref{sec:obs_red}. We update the host star properties using \textit{Gaia} DR2 and a new gyrochronological model (Section \ref{sec:star_props}). The RV measurements (Section \ref{sec:orbit_analysis}) allow us to provide an improved mass estimate and to detect a second planet in the system (Section \ref{sec:res}). The new mass estimate constrains WASP-107b's envelope fraction and allow us to propose a possible explanation for the origin of close-in exo-Neptunes like WASP-107b (Section \ref{sec:disc}).

\section{Spectroscopic Observations}\label{sec:obs_red}

\subsection{Keck/HIRES}

We collected a total of 60 radial velocity measurements of WASP-107 over 56 nights between 2017 and 2020 using the High Resolution Echelle Spectrometer~\citep[HIRES;][]{Vogt1994} on the Keck I Telescope. The observations were conducted by the California Planet Search (CPS; PI: Andrew Howard). The bulk of the observations was obtained as part of a K2-NASA key project between 2017 and early 2018. Following this program, we began re-monitoring the system in 2019 to better constrain the orbit of a possible second planet. The data were acquired using the ``C2'' decker, integrating until the exposure meter reached 60,000 counts (S/N $\sim 100$/pixel, $\sim 15$ minutes). Wavelength calibrations were done using the iodine cell~\citep{Butler1996}. The HIRES data reduction followed the standard procedures of the California Planet Search described in~\citet{Howard2010}. Typical HIRES RV uncertainties are $\sim 1.5$~m/s. For two of the nights, in which we obtained three measurements, we bin the data in units of 0.1~days, yielding a total of 57 HIRES RVs.

\subsection{CORALIE}

We supplement our radial velocity dataset with 31 CORALIE measurements obtained between 2011 and 2014, as reported in \citet{Anderson2017}. The CORALIE measurements have a typical uncertainty of $\sim 10$~m/s (see Table \ref{tab:RV data}), 6 times larger than the typical uncertainties from HIRES, but improve our baseline. With the 31 CORALIE RVs, kindly provided to us by D. R. Anderson, our combined dataset consists of 91 RVs spanning $\sim$ 3400~days, with a 1072~day gap between the two (Figure \ref{fig:multipanel}). These RVs are included in the machine-readable version of Table \ref{tab:RV data}. 
\begin{deluxetable}{lrrc}
\centering
\tablecaption{Sample radial velocities of WASP-107}
\tablehead{
  \colhead{Time} & 
  \colhead{RV} & 
  \colhead{$\sigma_\mathrm{RV}$} & 
  \colhead{Instrument} \\
   \colhead{(BJD$_{\mathrm{TDB}}$)} & 
  \colhead{(m s$^{-1}$)} & 
  \colhead{(m s$^{-1}$)} & 
  \colhead{}
}
\startdata
2457760.081152 & -1.38 & 1.72 & HIRES  \\
2457760.121746 & 4.83 & 1.57 & HIRES  \\
2457760.164031 & 7.16 & 1.55 & HIRES  \\
2457764.124912 & -13.31 & 1.55 & HIRES  \\
2457765.058488 & -2.94 & 1.50 & HIRES  \\
2457766.062488 & 3.33 & 1.51 & HIRES  \\
2457788.098712 & 1.62 & 1.72 & HIRES  \\
2457789.081827 & 3.72 & 1.71 & HIRES  \\
2457790.082269 & -4.18 & 1.57 & HIRES  \\
2457791.002344 & -16.74 & 1.61 & HIRES  \\
2457793.072951 & -5.30 & 1.48 & HIRES  \\
2457794.970762 & 12.43 & 1.69 & HIRES  \\
2457802.923643 & -27.41 & 1.46 & HIRES  \\
... & ... & ... & ...\\
2455651.762169 & -7.43 & 9.87 & CORALIE  \\
2455674.666370 & -9.29 & 7.05 & CORALIE  \\
2455683.663183 & 8.62 & 6.93 & CORALIE  \\
2455706.521724 & 27.38 & 7.72 & CORALIE  \\
2455763.493443 & 8.98 & 7.96 & CORALIE  \\
2455963.760279 & 37.27 & 9.75 & CORALIE  \\
2455983.823533 & -19.14 & 7.17 & CORALIE  \\
2456069.631011 & -21.18 & 10.55 & CORALIE  \\
... & ... & ... & ...
\enddata
\tablecomments{A full version, including activity indicators, is available as a machine readable table in the electronic journal.}
\label{tab:RV data}
\end{deluxetable}

\section{Host Star Properties} \label{sec:star_props}

The precision and accuracy of exoplanet parameters are tied to that of the ones of its host star. Here we refine the stellar parameters using our new Keck/HIRES spectra and derive a new estimate for the stellar age using gyrochronology.

\subsection{Updated Stellar parameters}

We use \verb|isoclassify| \citep{Huber2017} in order to provide updated estimates for the stellar radius, mass, age and luminosity. We adopt as inputs to \verb|isoclassify| the $\mathrm{T}_{\mathrm{eff}}$ and [Fe/H] values from the Keck/HIRES spectra, constrained by the \verb|SpecMatch-emp| tool \citep{Yee2017}. \verb|SpecMatch-emp| provides constraints on the the stellar parameters by comparing the Keck/HIRES spectra with an empirical spectral library. We use \textit{JHK} 2MASS magnitudes, while the parallax, right-ascension, and declination are taken from \textit{Gaia} DR2 (see Table \ref{tab:starParam}). 

Our reported value of $\log~g$ in Table \ref{tab:starParam} was obtained using the grid-modeling method in \verb|isoclassify| with a loose input constraint on $\log g$ ($4.7 \pm 0.2$). The constraint on $\log g$ that we obtain from this first fit is used as an input to a second fit using the direct method. 

The parameters that were derived by both the direct and grid-modeling methods (stellar radius and luminosity) are consistent within less than 1$\sigma$. We report the radius and luminosity determined using the direct method, as in this mode the radius is not affected by the uncertainty on stellar models. The values that we quote for the stellar mass and age are the results obtained using the grid-modeling method. 

The values of all parameters yielded by \verb|isoclassify| are consistent with \citet{Anderson2017} and \citet{Mocnik2017}, and our mass estimate is more precise, which we attribute to the higher precision of the $\mathrm{T}_{\mathrm{eff}}$ and [Fe/H] obtained from our HIRES spectra.

The updated stellar parameters are listed in Table \ref{tab:starParam}. We use the new stellar radius in order to update the radius and semi-major axis of WASP-107b; taking $R_b/R_\star$ and $a_b/R_\star$ from \citet{DaiWinn2017}, we find $R_b=0.96 \pm 0.03 R_J$ and $a_b=0.0566 \pm 0.0017$ AU.

\subsection{Stellar age}

The age of WASP-107 was previously estimated using several methods: an isochronal age estimate using {\sc bagemass} \citep{Maxted2015} yielded $8.3 \pm 4.3$ Gyr -- in agreement with the constraint we obtain with \verb|isoclassify|, and a conflicting gyrochronological age of $0.6 \pm 0.2$ Gyr \citep{Mocnik2017}.
This discrepancy can be explained by the failure of standard spin-down models to reproduce observations of reduced braking efficiency for late-K dwarfs relative to their F and G-type counterparts, leading them to experience a period of stalled rotation periods past $\sim 700$ Myr \citep{Agueros2018,Curtis2019,Curtis2020,Curtis2020b}.

To resolve the discrepancy, we use a new gyrochronology model that accounts for stalled magnetic braking (Angus et al., in prep) in order to estimate the age of WASP-107. This model was calibrated by fitting a Gaussian process -- a semi-parametric model that is flexible enough to capture the complex nature of stellar spin-down, to a number of asteroseismic stars and open clusters, including NGC 6811 which exhibits stalled magnetic braking \citep{Curtis2019}. We also use kinematic ages of {\it Kepler} field stars to calibrate the model for old K and early M dwarfs, where there is a dearth of suitable open cluster calibration stars. These kinematic ages also reflect the stalled magnetic braking behaviour seen in open clusters \citep{Angus2020}.

Using our new gyrochronology model, we infer an age of 3.4 $\pm$ 0.3 Gyr for this star using the measured rotation period of $17 \pm 1$ days \citep{Anderson2017}. The 0.3 Gyr uncertainty is the formal uncertainty that results from the uncertainty on the star's rotation period, and does not account for uncertainty in the model. Quantifying the magnitude of the model uncertainty is beyond the scope of this paper and we adopt a 20\% value of 0.7 Gyr as a more reasonable estimate of the true age uncertainty (Table \ref{tab:starParam}). Importantly, we resolve the conflict betwen the age estimates and conclude that WASP-107 is not as young as previously suggested by gyrochronology models.

\begin{table*}
\centering
\begin{tabular}{lll}
\hline
\hline
 Parameter  &Value &Source\\
\hline
\textit{~~~~Identifying information}&&\\
EPIC ID & 228724232  & \\
$\alpha_{J2000}$ R.A. (hh:mm:ss)&12:33:32.84 & \textit{Gaia} DR2 \citep{Gaia2018}\\
$\delta_{J2000}$ Decl. (dd:mm:ss)&-10:08:46.22 & \textit{Gaia} DR2 \citep{Gaia2018}\\
\textit{~~~~Photometric properties}&&\\
B (mag) & 14.62$\pm$1.20& \citet{Hog2000}\\
V (mag) & 11.47$\pm$0.20& \citet{Hog2000}\\
G (mag) & 11.1740$\pm$0.0009& \textit{Gaia} DR2 \citep{Gaia2018}\\
J (mag) & 9.378$\pm$0.021 & 2MASS \citep{Cutri2003}\\
H (mag) & 8.777$\pm$0.026 & 2MASS \citep{Cutri2003}\\
K (mag) & 8.637$\pm$0.023 & 2MASS \citep{Cutri2003}\\
\textit{~~~~Spectroscopic and Derived properties}&&\\
$\mu_\alpha$ (mas yr$^{-1}$) & -96.647$\pm$0.110 & \textit{Gaia} DR2 \citep{Gaia2018}\\
$\mu_\delta$ (mas yr$^{-1}$) & -9.483$\pm$0.058 & \textit{Gaia} DR2 \citep{Gaia2018}\\
Parallax (mas) &  15.4175$\pm$0.0617 & \textit{Gaia} DR2 \citep{Gaia2018}\\
Distance (pc) & 64.7$\pm$0.3 & \citet{Bailer2018}\\
Spectral Type & K6 & \citet{Anderson2017}\\
$\mathrm{T}_{\mathrm{eff}}$ (K) & $4425 \pm 70$ & This Paper (\verb|SpecMatch-emp|)\\
$\left[ Fe/H \right]$ & $+0.02\pm 0.09$ & This Paper (\verb|SpecMatch-emp|)\\
$log~g_\star$ & 4.633$\pm 0.012$ & This Paper (\verb|isoclassify|)\\ 
$M_\star$  ($M_\odot$) &0.683$^{+0.017}_{-0.016}$ & This Paper (\verb|isoclassify|)\\
$R_\star$ ($R_\odot$) & 0.67$ \pm 0.02$ & This Paper (\verb|isoclassify|)\\
$L_\star$ ($L_\odot$) &  0.132$\pm$0.003 & This Paper (\verb|isoclassify|)\\
P$_{\mathrm{rot}}$ (days) & 17$\pm$1 & \citet{Anderson2017}\\
Age (Gyr) & 8.3$\pm$4.3 & \citet{Mocnik2017}; Isochronal ({\sc bagemass})\\
 &  6.9$^{+3.7}_{-3.4}$ & This Paper; Isochronal (\verb|isoclassify|)\\
 & 3.4$\pm$0.7 & This Paper; Gyrochronological\\
\hline

\end{tabular}
\caption{\label{tab:starParam}Stellar Parameters of WASP-107}
\end{table*}

\section{Analysis} \label{sec:orbit_analysis}

We use Keplerian model fitting and Bayesian model comparison  in order to robustly infer planetary parameters and search for additional planets in the system using the Keck/HIRES and CORALIE radial velocity observations.

\subsection{Radial-velocity analysis} \label{ssec:rvfit}

We fit the RV measurements using RadVel, an open source Python package for fitting Keplerian orbits to radial velocity datasets~\citep{radvel}. The posterior distributions of the parameters are sampled using the Markov Chain Monte Carlo (MCMC) affine-invariant sampler \texttt{emcee} \citep{ForemanMackey2013}. We fix the period $P_b$ and time of conjunction $T_{\mathrm{conj,b}}$ of WASP-107b to the \textit{K2} transit ephemeris reported in \citet{DaiWinn2017}. We use the conventional unbiased basis formed by the jump parameters $h = \sqrt{e} \cos \omega$ and $k = \sqrt{e} \sin \omega$ to fit for orbital eccentricities and arguments of periastron. The MCMC fitting basis is thus ($K_b$, $ \sqrt{e_b}\cos\omega_b$, $\sqrt{e_b}\sin\omega_b$), along with ($P_c$, $T_{\mathrm{conj,c}}$, $K_c$, $\sqrt{e_c}\cos\omega_c$, $\sqrt{e_c}\sin\omega_c$) for two-planet models. We introduce an offset term $\gamma_i$ and a jitter term $\sigma_i$ for each instrument ($i=\{H,C\}$ for HIRES and CORALIE). These terms account respectively for different zero-points between HIRES and CORALIE and for additional noise (e.g. stellar jitter) not encapsulated by the single-measurement errors. The offset terms are computed using a linearized analytic solution. We invoke physical priors on the RV semiamplitude $K > 0$ and eccentricity $e \in [0, 1)$. 
This brings the total of fitting parameters to 17 for a joint 2-planet fit of the HIRES and CORALIE data. 

We perform a maximum-a-posteriori (MAP) fit and use the best-fit parameters to seed a MCMC that estimates the full posterior. We run the MCMC with 4 ensembles of 200 walkers each of which are allowed to burn-in until the Gelman-Rubin statistic is $< 1.01$, checking for convergence after 2,000 steps per walker. We then save every 5$^{\mathrm{th}}$ step as a posterior sample until either 10,000 samples per walker are reached or the Gelman-Rubin statistic decreases below 1.005, at which point we consider the chains to be well-mixed. A second MAP fit is then run, starting at the median values determined by the MCMC posteriors. 

We also attempted to use Gaussian process (GP) regression in order to mitigate the effect of stellar activity on the RVs by training the GP on ancillary \textit{K2} time series, but found that adding the GP parameters to the RV fit was disfavored from a model comparison point of view (using Bayes factors; see Section \ref{ssec:model_comp}) and did not improve our constraints on the planet parameters. 

\subsection{Model Comparison} \label{ssec:model_comp}

We perform a total of four fits: a 1-planet and 2-planet fit to the HIRES RVs alone, and the same two fits to the combined HIRES and CORALIE dataset. Model comparison is performed within a robust Bayesian framework, using the Bayesian evidence and Bayes factors assuming equally probable hypotheses a-priori. A higher Bayesian evidence for a model including more free parameters indicates that the additional model complexity is granted by the quality of the data. Hence, the model with the highest Bayesian evidence is favored. For each model of the RVs, we approximate the Bayesian evidence using the estimator presented in \citet{Perrakis2014} (as in e.g. \citealp{Cloutier2019}). This method consists in using the product of marginalized posterior distributions of ``blocks'' of parameters (obtained after a rearrangement of the parameters in MCMC chains) as an importance sampler in the evidence estimation. \citet{Nelson2020} found that this estimator obtains results in good agreement with more computationally-intensive methods (e.g., Nested Sampling) while requiring no additional sampling of the parameter space. 
On the other hand, they found that non-bayesian methods such as Bayesian Information Criteria (BIC) that are based on simplifying assumptions such as an infinitely narrow posterior only poorly approximate the Bayesian evidence and can lead to false-negatives in planet detections using RVs. 
In this work, we use the implementation of this estimator in the \texttt{bayev} package, publicly-available on GitHub\footnote{\url{https://github.com/exord/bayev}}. Each block consists of only one of the fitted parameters. The importance sampler is obtained after a random rearrangement of the parameters and we use kernel density estimations of each marginalized posterior distribution of samples. For each model, we perform 1,000 such rearrangements, calculating the Bayesian evidence for each and quote the median from these values.

\section{Results}\label{sec:res}


\subsection{Detection of a long-period companion}

\begin{figure}
    \centering
    \includegraphics[width=0.47\textwidth]{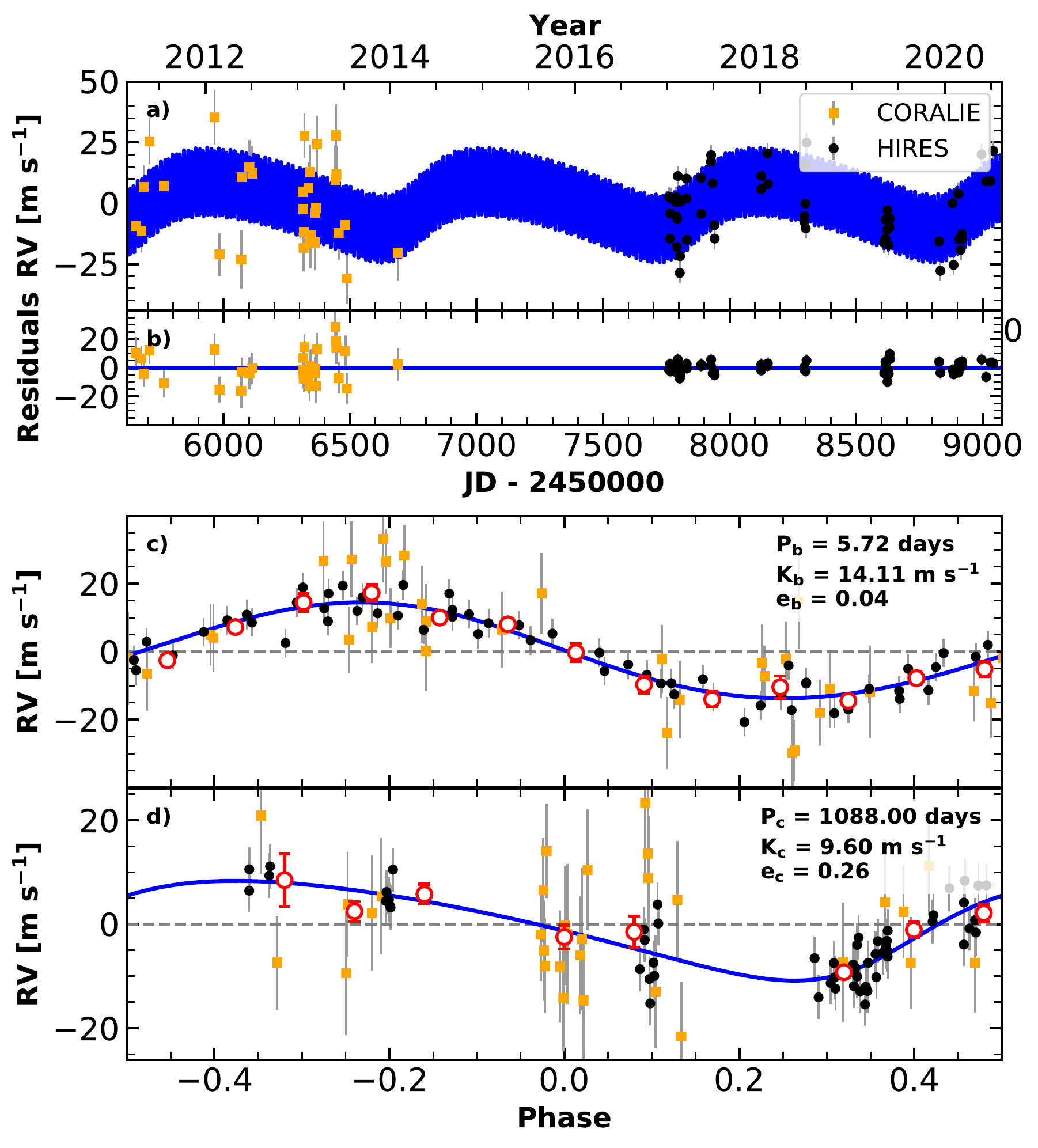}
    \caption{Maximum-likelihood two-planet Keplerian orbital model for WASP-107. {\bf a)} The best-fitting model is plotted while the annotated orbital parameters are the corresponding MAP fit parameters. The blue line is the best fit two-planet model. We add in quadrature the RV jitter terms listed in Table~\ref{tab:params} with the measurement uncertainties for all RVs.  {\bf b)} Residuals to the best fit two-planet model. {\bf c)} RVs phase-folded to the ephemeris of planet b. The Keplerian orbital model for the other planet has been subtracted. The small point colors and symbols are the same as in panel {\bf a}.  Red circles are the same velocities binned in units of 0.08 in orbital phase.  The phase-folded model for planet b is shown as the blue line, with the corresponding best-fit keplerian model parameters quoted. {\bf d)} The same for planet c.}
    \label{fig:multipanel}
\end{figure}

The HIRES data exhibit a significant long-period trend on top of the signal from WASP-107b due to the presence of a second planet in a long period orbit (Figure \ref{fig:multipanel}). Our Bayesian model comparison based on the HIRES data alone favors the two-planet model at $4.9 \times 10^7$ to 1 in the Bayes factor (equivalently 6.3$\sigma$; Table \ref{tab:Bayesian comparison}). The evidence for the existence of this second planet, WASP-107c, is further strengthened by including the previously obtained CORALIE dataset in the radial velocity analysis (see Table \ref{tab:Bayesian comparison}). Overall, the CORALIE dataset increases the baseline by a factor of $\sim 3$, allowing us to rule out a 1-planet model in favor of a two-planet Keplerian solution, with a Bayes factor of $1.2 \times 10^{58}$ to 1 (or a 16.5$\sigma$ significance). The best-fit two-planet model to the combined dataset is shown in Figure~\ref{fig:multipanel}. In what follows, we will report constraints on the orbital parameters based on the two-planet model and the combined dataset.

\begin{deluxetable}{lccc} \label{tab:Bayesian comparison}
\centering
\tablecaption{Keplerian model comparison. The Bayes factors are reported relative to the single-planet model. In both cases, the 2-planet model is favored.}
\tablehead{
  \colhead{Model} & 
\colhead{$\ln p(D|\mathcal{M}_i)$} &
\colhead{$\frac{p(D|\mathcal{M}_2)}{p(D|\mathcal{M}_1)}$ } &
\colhead{``Sigma''\tablenotemark{a}}
}
\startdata
\sidehead{\bf{HIRES}}
1-Planet               & -174.3 & -- & -- \\
2-Planet              & -156.6 & $4.9 \times 10^{7}$  & 6.3$\sigma$\\
\hline
\sidehead{\bf{HIRES + CORALIE}}
1-Planet                & -290.1 & -- & --\\
2-Planet               & -156.4 & $1.2 \times 10^{58}$ & 16.5$\sigma$
\enddata
\tablenotetext{\tiny a}{The correspondence between the Bayes factors and the ``sigma''  significance in a frequentist framework is calculated using p-values, as presented in \citet{Trotta2008}.}
\end{deluxetable}

\begin{deluxetable}{lcc}
\centering
\tablecaption{MCMC Posteriors for the WASP-107 System}
\tablehead{
  \colhead{Parameter} & 
  \colhead{Unit} & 
  \colhead{Credible Interval}
}
\startdata
\sidehead{\bf{Orbital Parameters}}
$P_b$ &days&    $\equiv 5.7214742$ \\
$T_{\rm{conj},b}$ & BJD$_\mathrm{TDB}$ &    $\equiv 2457584.329897$  \\
$e_b$  &&   $0.06 \pm 0.04$\\ 
$\omega_b$ &deg &  $40^{+40} _{-60}$\\
$K_b$& m s$^{-1}$  & $14.1 \pm 0.8$ \\
\\
$P_c$ & days & $1088 ^{+15 } _{-16 }$ \\
 & yrs & $2.98 \pm 0.04$ \\
$T_{\rm{conj},c}$ & BJD$_\mathrm{TDB}$ &  $2458520 ^{+ 60} _{- 70 }$ \\
$e_c$  &  & $0.28 \pm 0.07$ \\
$\omega_c$ & deg &  $-120 ^{+ 30 } _{-  20 }$ \\
$K_c$ & m s$^{-1}$  &  $9.6 ^{+1.1 } _{- 1.0 }$\\
\sidehead{\bf{Derived Parameters}}
$M_b$ &M$_\oplus$ &    $30.5 \pm 1.7$ \\
 & M$_J$ &  $0.096 \pm 0.005$\\
$\rho_b$ & g cm$^{-3}$ & $0.134 ^{+ 0.015 } _{-  0.013 }$ \\
\\
$M_c~ \sin i$ & M$_\oplus$ &  $115 \pm 13$ \\
 & M$_J$ &  $0.36 \pm 0.04$ \\
$S'_c$\tablenotemark{a} &  mas & $26_{-5}^{+8}$\\
\sidehead{\bf{Global Parameters}}
$\gamma_\mathrm{H}$\tablenotemark{b} & m s$^{-1}$ &   $\equiv 1.52001$\\
$\gamma_\mathrm{C}$\tablenotemark{b}& m s$^{-1}$ &   $\equiv 3.55333$ \\
$\sigma_\mathrm{H}$ & m s$^{-1}$ &  $3.9 ^{ +0.5 } _{  -0.4 }$ \\
$\sigma_\mathrm{C}$ &m s$^{-1}$ &   $5.5^{+2.8 } _{ -2.9 }$
\enddata
\tablenotetext{\tiny a}{Sky-projected angular separation}
\tablenotetext{\tiny b}{
Reference epoch for $\gamma$: 2457934.793533}
\label{tab:params}
\end{deluxetable}

\begin{figure*}
    \centering
    \includegraphics[width=\textwidth]{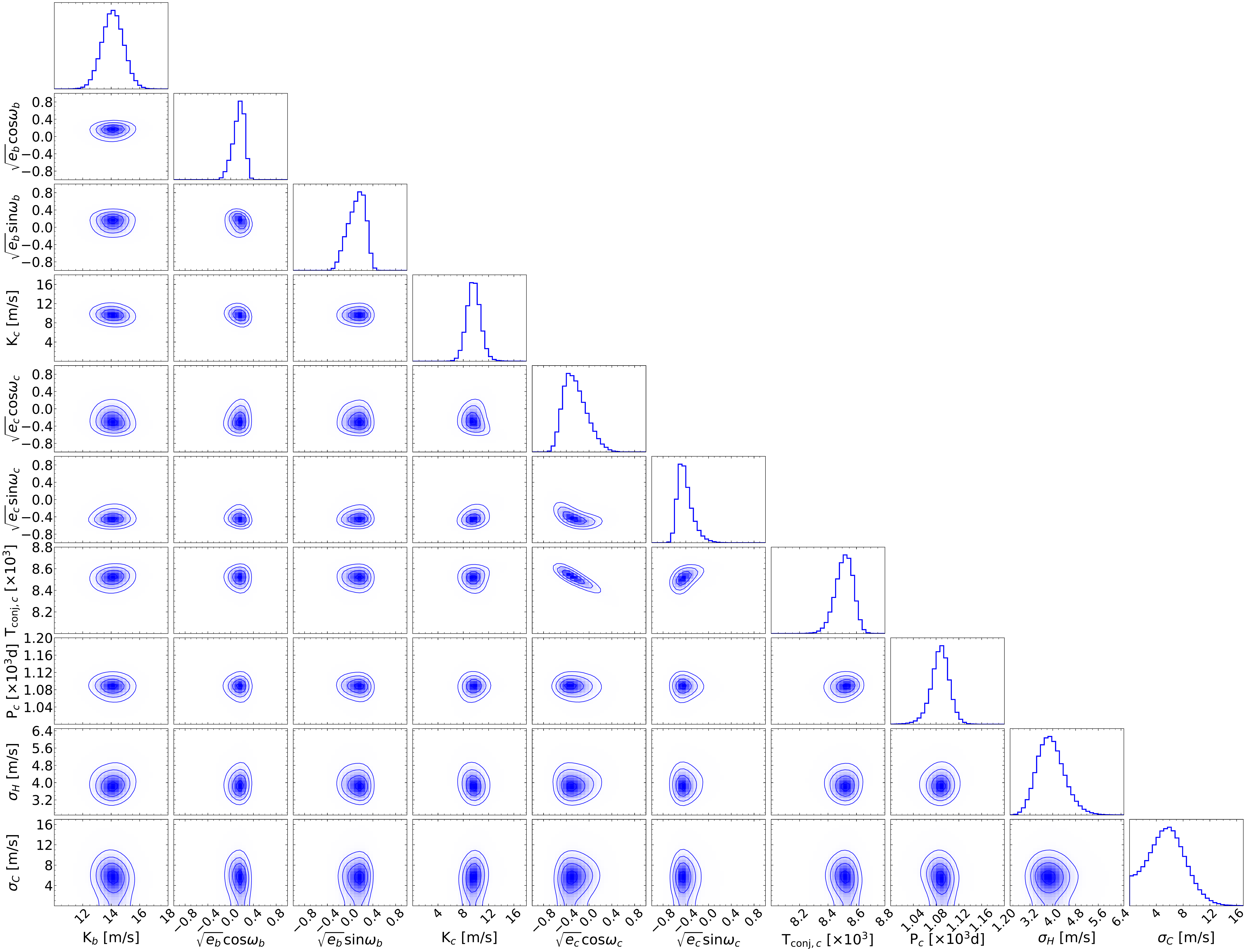}
    \caption{Joint and marginalized posterior distributions from the two-planet fit to the combined HIRES+CORALIE dataset.}
    \label{fig:RV_corner}
\end{figure*}

\subsection{Mass and Orbital Constraints} \label{ssec:mass_orb_res}

\begin{figure*}
\centering
\includegraphics[width=0.8\linewidth]{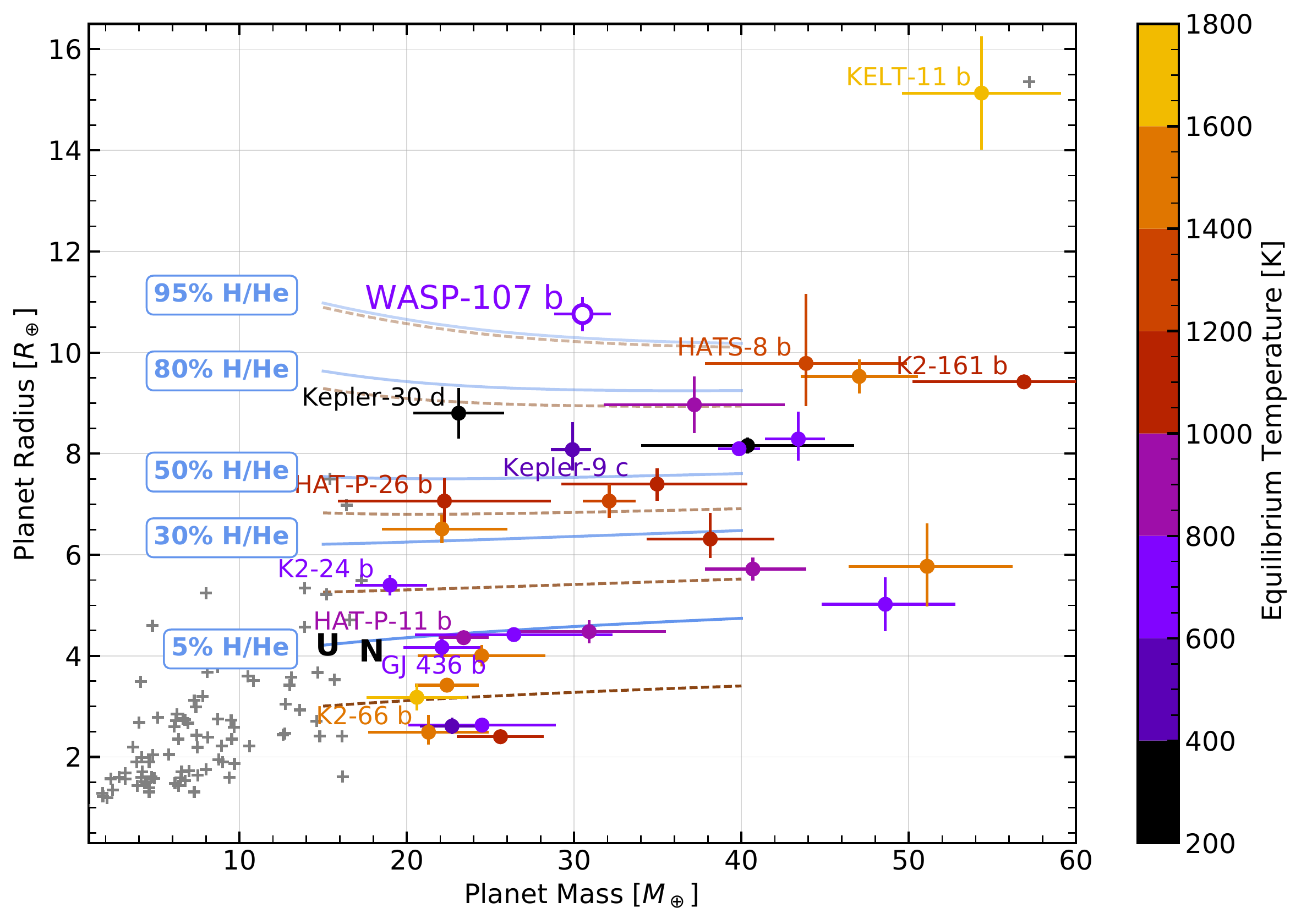}
\caption{\label{fig:env_frac} Mass-radius diagram of the detected exoplanets with a mass and radius measured with 30\% or higher precision, zoomed-in on super-Neptunes having masses between 20 and 60 M$_\earth$ (colored by equilibrium temperature). The positions of Uranus and Neptune are indicated by black letters. The solid blue and dashed saddlebrown curves correspond to various envelope mass fractions from the grid of solar-metallicity envelopes atop pure H$_2$O or Earth-like cores respectively, interpolated to match an age of 3.4 Gyr and the incident flux corresponding to WASP-107b. If no inflation mechanism is considered, we obtain a 3$\sigma$ lower limit on the envelope mass fraction of 85\%, which corresponds to a core mass of only 4.6 $M_\oplus$. The larger KELT-11b may be inflated due to its high equilibrium temperature \citep{Pepper2017}.}
\end{figure*}

MCMC parameter estimates and uncertainties are displayed in Table~\ref{tab:params} along with derived planet parameters. The corresponding joint and marginalized posterior distributions  are shown in Figure \ref{fig:RV_corner}. We find a revised mass for WASP-107b of $M_b = 30.5 \pm 1.7$~$M_\oplus$ and an eccentricity of $e_b = 0.06 \pm 0.04$. This lower mass associated with a Jupiter radius makes WASP-107b an outlier in the mass-radius diagram (Figure \ref{fig:env_frac}). Combined with the transit measurement for the radius, we find the bulk density of WASP-107b to be $\rho_b = 0.134^{+0.015}_{-0.013}$~g~cm$^{-3}$, even lower than the previous estimate of $0.19 \pm 0.03$~g~cm$^{-3}$ \citep{Anderson2017}. 

The HIRES and CORALIE datasets combined constrain the RV semi-amplitude of the outer companion to $K_c = 9.6^{+1.1}_{-1.0}$ m~s$^{-1}$, corresponding to a mass of $M_c~\sin~i= 0.36 \pm 0.04$~M$_{J}$. The orbital period of the outer planet ($2.98 \pm 0.04$ years) and its eccentricity ($e_c = 0.28 \pm 0.07$) are well constrained thanks to the detection of two steep rises in RV throughout the HIRES observing campaign.
Continued monitoring of this system will enable further refinements of the orbit of this outer planetary companion.

\subsection{An even lower mass for WASP-107b}

Our measurement of WASP-107b's mass of $30.5 \pm 1.7~M_{\oplus}$ is substantially lower than the inferred $38\pm3~M_\oplus$ reported by \citet{Anderson2017} from the analysis of the CORALIE dataset. We argue that this discrepancy can be explained as arising from the improved precision of the HIRES RV measurements.

Our new mass estimate for WASP-107b is 20\% lower than the previously-published one, and lies $\approx 2.2\sigma$ away from it. For a fitted semi-amplitude $K$ to a time series of $N_{obs}$ radial velocities having a typical measurement uncertainty $\sigma$, $K$ will be biased towards a higher value than the true one if $K/\sigma$ or $N_{obs}$ is small \citep{Shen2008}. Thus, with $N_{obs}=31$ and $K/\sigma<2$, the CORALIE dataset is expected to suffer from this bias more than the HIRES measurements, where $N_{obs}=60$ and $K/\sigma \approx 11$. Moreover, \citet{Anderson2017} fixed the eccentricity to zero (yielding smaller uncertainties on $K$) and used a larger stellar mass compared to our new determination, which further contributed to a larger estimated planet mass. 

\subsection{Gas-to-core mass ratio}\label{ssec:gcr}

Our new mass measurement indicates that WASP-107b has an extraordinarily low density. We infer the gas-to-core mass ratio of WASP-107b by comparing our measurements to model predictions for various envelope mass fractions (Figure \ref{fig:env_frac}). 
We generate grids of thermal evolution models and predict 10-mbar radii for planets with varying incident fluxes $S_{inc}$, masses $M_p$, ages and fractions $f_{env} [\%]$ of their masses contained in the H/He envelopes surrounding their cores. Previously available model grids in the literature (e.g. \citealp{Lopez2014}) do not extend all the way to the mass and radius of WASP-107b and could not yet make use of the latest equation of state for hydrogen.

We compute the thermal evolution models as described in \citet{Thorngren2016}. The atmosphere models are interpolated from the \citet{Fortney2007} solar-metallicity grids and serve as a boundary condition to interior structure models. We use updated equations of state (EOS) for the solar-ratio H/He gas \citep{Chabrier2019} and choose the \citet{Thompson90} EOS for metals. Since the composition of the core of WASP-107b is unknown, we marginalize over the uncertainty on the fraction of water within its core by producing two grids for planets with either an Earth-like core (67.5\% rock, 32.5\% Fe by mass) or a pure water composition. Therefore, the true envelope mass fraction of the planet should lie in the range cornered by the two derived envelope mass fractions: the same planet radius and mass will yield a larger inferred H/He mass fraction if the core composition is rocky than if it is water-rich. The two-layer models include no mixing from the core into the envelope and in the pure-water core case, we include convection in the water component. Our grids span the range 10 to 10$^5$ $S_\oplus$ for incident flux, 15 to 40 $M_\oplus$ for planet mass, 0.1 to 14 Gyr in terms of system age and H/He envelope mass fractions of 0.01 to 100\%. The curves of constant $f_{env}$ in the mass-radius diagram computed using the median parameters for WASP-107b already suggest an extremely high envelope mass (Figure \ref{fig:env_frac}).

More quantitatively, we solve the inverse problem of inferring an envelope mass fraction from the planet's incident flux, mass, age and radius using the \texttt{smint} (Structure Model INTerpolator) interpolation and envelope mass fraction fitting package, which we made publicly-available on GitHub\footnote{\url{https://github.com/cpiaulet/smint}}. 
From any set of parameters ($S_{inc}$, $M_p$, age, $f_{env}$), \texttt{smint} returns a planet radius by performing linear interpolation over a grid of $f_{env}$, $\log_{10} M_p/M_\oplus$, system age and $\log_{10} S_{inc}/S_\oplus$ set up using one of our model grids. 
We run a MCMC that fits for the combination of ($S_{inc}$, $M_p$, age, $f_{env}$) which best matches the observed planet radius. We adopt gaussian priors on $S_{inc}$, $M_p$ and the system's age informed by the parameters derived from the RV fits and the gyrochronology model. We use a uniform prior on the planet's envelope mass fraction over the entire range spanned by the grid. 
Each of the 100 chains is run for 10,000 steps, 60\% of which are discarded as burn-in. The autocorrelation times of the fitted parameters are all $\leq 52$ steps (less than 15\% of the length of the chains past the burn-in phase), securing that our chains are converged and sample well the posterior PDFs. We perform two fits using the grids for the Earth-like and pure H$_2$O core compositions.

\begin{figure}
\centering
\includegraphics[width=0.99\linewidth]{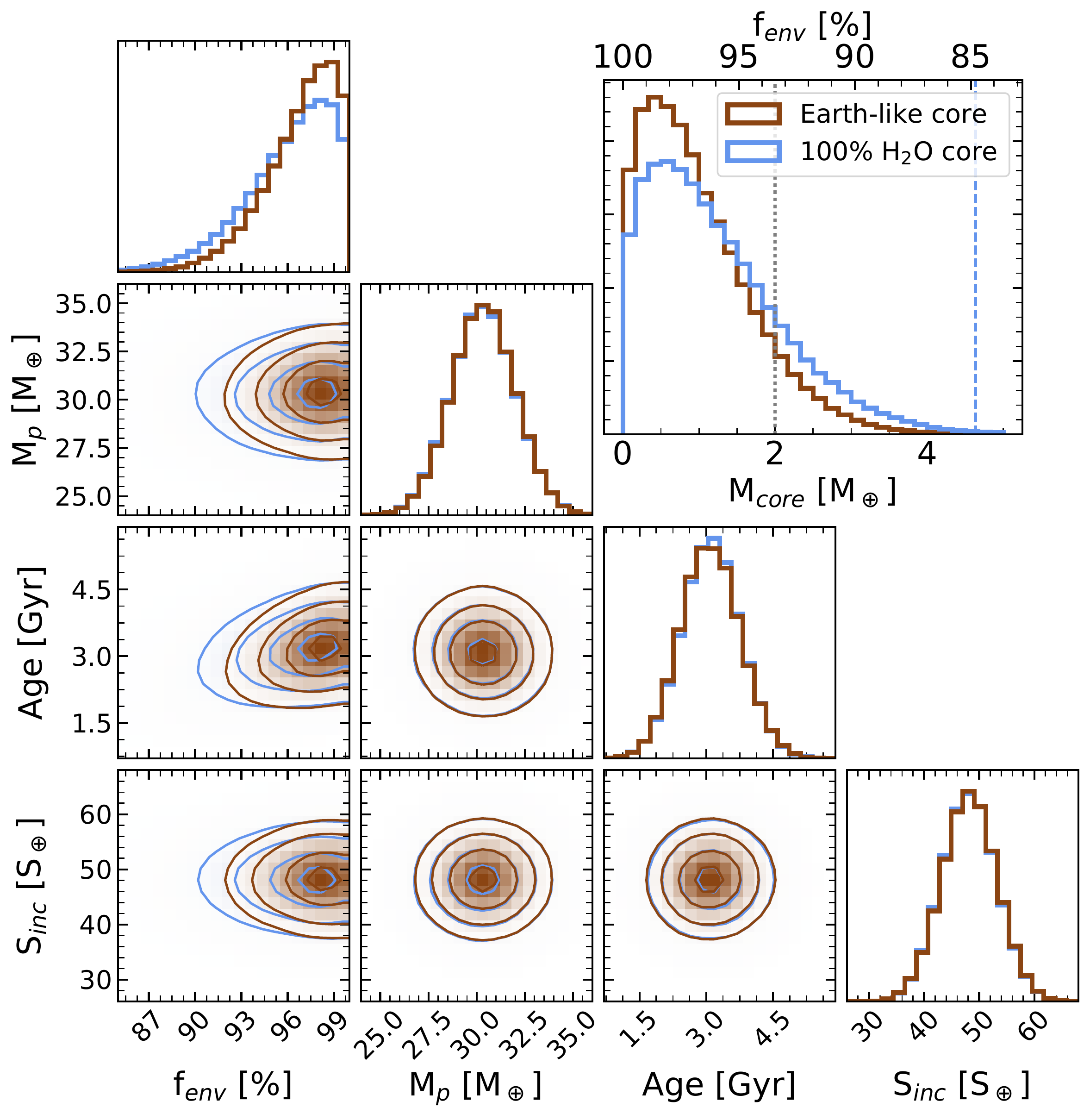}
\caption{\label{fig:corner_fenv} Constraints on the envelope mass fraction and core mass of WASP-107b from our models assuming either an Earth-like or a pure-water core compostion. The joint and marginalized posterior distributions for a 1$\times$ solar metallicity envelope are shown in blue and gray respectively, along with the corresponding distribution of core masses. The dotted gray line corresponds to a core mass of 2 $M_\oplus$ which we use in our envelope accretion simulations (Section \ref{sec:disc}). We place a 3$\sigma$ upper limit of 4.6 $M_\oplus$ on the mass of the pure water core (dashed blue line).}
\end{figure}

We infer upper limits on the core mass of $3.7$ and $4.6$ $M_\oplus$ at 3$\sigma$ (lower masses are excluded with 99.7\% confidence) for the Earth-like and pure H$_2$O core respectively, corresponding to envelope mass fractions of $88$\% and $85$\% for a solar metallicity envelope (Figure \ref{fig:corner_fenv}). 
As expected, we find a correlation between the system's age and the inferred envelope mass fraction due to the contraction and cooling of the planet as it ages. The inferred core mass is much smaller than traditionally assumed to be required for the triggering of runaway gas accretion. Interestingly, the small core still must have been able to accrete almost 30~$M_\oplus$ in gas from the surrounding nebula once it had coagulated.

Our estimate also serves as a conservative lower limit on the initial accreted envelope mass, given that planets undergo atmospheric loss throughout their existence (e.g. \citealp{Owen2013,Owen2016}). In fact, the puffiness and close-in orbital distance of WASP-107b places it at the cusp of the sub-Jovian desert (e.g. \citealp{Mazeh2016,Owen2018}). 
In particular, we find that assuming a zero Bond albedo, WASP-107b's core would need to be more massive than 4.1 $M_\oplus$ (1$\sigma$ upper limit accounting for the uncertainties on the stellar and orbital parameters) for the atmosphere to survive photoevaporation at its present orbital location \citep{Ginzburg2016}. This limit is higher than our above-mentioned estimates of the core mass, suggesting that WASP-107b migrated to its present orbital location only recently (see Section \ref{ssec:dyn}).

The large H/He mass fraction of WASP-107b strikingly contrasts with those of the ice giants of our Solar System (5--15\% H/He for both Uranus and Neptune, \citealp{Podolak91,Guillot2005}). Instead, WASP-107b's structure most closely resembles those of Jupiter and Saturn, composed of more than 90\% H/He by mass. WASP-107b is an extreme case even among the vastly diverse exoplanet interiors illustrated by the wide span in radius of the super-Neptunes, which highlights the great diversity of formation pathways within this population. WASP-107b, with its Jupiter-like composition, resembles K2-161b and Kelt-11 b that have large envelope mass fractions ($>75$\%) associated with their low densities \citep{Brahm2019,Pepper2017}. Meanwhile, thermal evolution models predict for instance a Neptune-like structure for the 23 $M_\oplus$ planet HAT-P-11b, with about 15\% H/He by mass \citep{Petigura2017}.

\begin{figure}
    \centering
    \includegraphics[width=0.95\linewidth]{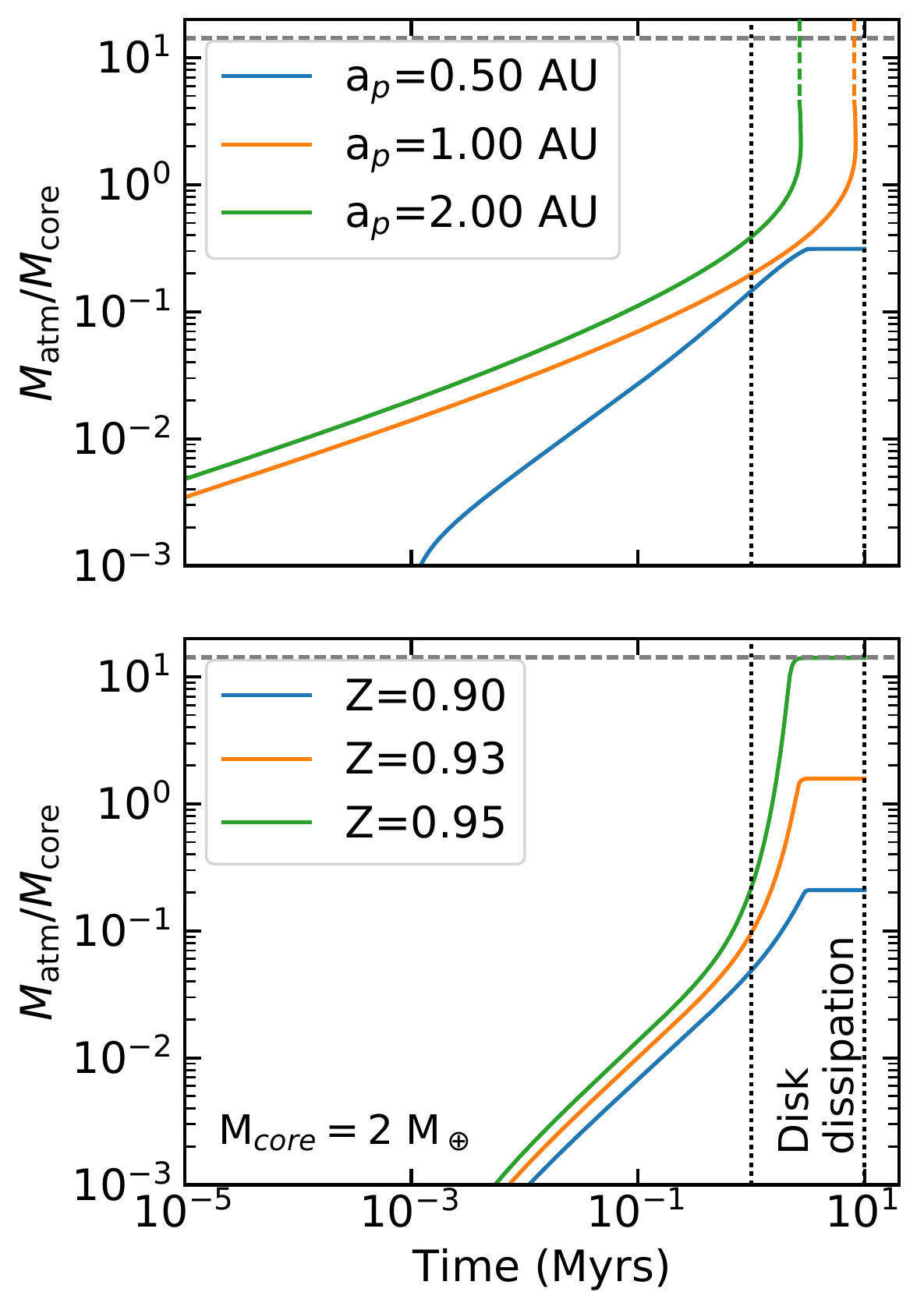}
    \caption{Potential formation pathways of WASP-107b for a 2 $M_\oplus$ core. Two different modes of gas accretion are explored: dust-free, solar metallicity gas (top panel) and dusty, metal-enriched gas (bottom panel). The orbital distance is freely varied in dust-free calculation while the dusty calculation was performed in situ (0.05 AU). The GCR that is consistent with the measured mass of WASP-107b for a 2 $M_\oplus$ core (dashed gray line) is above unity, well beyond the triggering of runaway, super-exponential growth. This simulation highlights the  plausible parameter space that can match the target GCR within the typical disk gas dissipation time (delimited by the vertical dotted lines). Inside 1 AU, runaway gas accretion can be quenched by the rapid disk gas dispersal before the planet blows up into a gas giant.}
    \label{fig:GCR_sim}
\end{figure}

\section{Implications for planet formation}\label{sec:disc}

In light of the results from our radial-velocity analysis, we find that WASP-107b likely has an extremely large gas-to-core mass ratio (GCR; Section \ref{ssec:gcr}). This makes it one of the puffiest
Neptune-mass planets known to date and raises two important questions: \textit{How could a core of $< 4.6$ $M_\oplus$ accrete such a massive envelope?} and \textit{How was its transformation into a Jupiter-mass gas giant thwarted?}
As such an extreme case, WASP-107b motivates a detailed investigation of the physics of gas accretion and orbital migration at play.

\subsection{Planet formation scenarios}

According to the core-nucleated accretion theory \citep{Perri74,Mizuno80,Pollack1996}, which stands as the ``standard model'' for planet formation, a planetary core will first assemble from the solids present in its neighborhood within the protoplanetary disk. Past this phase of solid accretion, a second phase will start, dominated by gas accretion. The rate of envelope accretion is governed by
cooling of the gas \citep{Lee2015,Ginzburg2016} and by local and global hydrodynamic flows \citep{Tanigawa2016}. \citet{Lee2019} demonstrated that the observed distribution of GCRs can effectively be reproduced if all these mechanisms are taken into account. For core masses below $\approx$ 20 $M_\oplus$, the cooling timescale remains the longest until the runaway is triggered (Figure 2 of \citealp{Lee2019}) beyond which the rapid dispersal of background disk limits the delivery of gas onto planets. We note here that this rapid dispersal is valid only within $\sim$1 AU where the photoevaporative wind is launched, carving out a gap disconnecting the inner disk from the outer disk \citep{Owen2011}.

To reproduce the inferred GCR of WASP-107b, we need to identify the gas accretion scenario in which the runaway growth begins but is curbed prematurely within the typical disk lifetime $\sim$1--10 Myrs \citep[e.g.,][]{Haisch2001,Mamajek2009,Pfalzner2014,Alexander2014}. We explore two potential pathways, motivated by the results of \citet{Lee2015}. First, we consider dust-free accretion at distances beyond the present-day orbit of WASP-107b. Second, we consider in-situ (0.05 AU) accretion of dusty, metal-enriched gas.

\subsection{Dust-free accretion with solar metallicity}

As dust grains coagulate and sediment, they no longer contribute to the atmospheric opacity \citep[e.g.,][]{Ormel2014,Mordasini2014,Piso2015}. The upper radiative layer becomes largely isothermal so that the envelope becomes colder at larger stellocentric distances. As the internal modes of gaseous molecules gradually freeze out, the opacity drops and the envelope cools faster and therefore grows faster. Dust-free accretion beyond $\sim$1 AU was invoked as a way to explain Kepler-51b, a super-puff of a $\sim$2 $M_\oplus$ core that accreted a $\sim$30\% by mass envelope \citep{Lee2016}.

For WASP-107b, we adopt the procedure of \citet{Lee2019}---where at each timestep, the rate of gas accretion is set by the minimum between cooling, local hydrodynamic, and global disk accretion---but replace the cooling calculation to that of dust-free accretion. One-dimensional structure equations are solved numerically following \citet{Lee2014}, corrected for the decrease in the bound radius by three-dimensional advective flows \citep[e.g.,][]{Lambrechts2017}. We confirm that the updated numerical solutions computed at a few orbital distances agree well with the analytic scaling relationship derived by \citet{Lee2015}, and search for the range of potential formation locations by scaling with respect to the fiducial numerical calculation at 2 AU (see top row of Figure \ref{fig:GCR_sim}). For a core that weighs 2 $M_\oplus$, dust-free accretion should have occurred at $\gtrsim 1$ AU for the runaway to be launched.

\subsection{Dusty accretion with supersolar metallicity}

Are there ways to form WASP-107b in-situ? It appears impossible for solar metallicity gas but the rate of accretion can be boosted in highly metal-enriched environments. While this may be counter-intuitive, as higher opacity would slow down the cooling of gas, this impediment by opacity endures only up to $Z \sim 0.2$ (10 times the solar value). Beyond this point, the mean molecular weight of the envelope becomes so high that it becomes more prone to gravitational collapse requiring more vigorous accretion in order to maintain hydrostatic equilibrium. This boost in accretion rate applies equally to dust-free and dusty accretion. Since dusty accretion is generally slower than dust-free accretion, we elect to show the former to delineate all the possible limiting cases.

We proceed the same way as in the calculation for dust-free accretion and follow the procedure of \citet{Lee2019}, accounting for the effect of metallicity in the rate of cooling, which can be written as
\begin{align}
    \frac{M_{\rm atm}}{M_{\rm core}} &\sim 0.09 \left(\frac{\Sigma_{\rm gas}}{13\,{\rm g\,cm}^{-3}}\right)^{0.12} \left(\frac{M_{\rm core}}{20\,M_\oplus}\right)^{1.7}\left(\frac{t}{0.1\,{\rm Myrs}}\right)^{0.4} \nonumber \\
    &\times \left(\frac{0.02}{Z}\right) \left(\frac{\mu}{2.37}\right)^{3.4} {\rm Exp}\left(\frac{t}{t_{\rm run}}\right)
\end{align}
where $\Sigma_{\rm gas}$ is the disk gas surface density evaluated at the position of the core and at time $t$,\footnote{The disk is evolved according to steady-state viscous dissipation \citep{Lynden-Bell1974,Hartmann1998} up to 0.7 Myrs. Beyond that, the inner disk detaches from the outer disk and viscously drifts in over the dissipation timescale evaluated at its outer edge \citep{Owen2011}.} $\mu$ is the mean molecular weight, and $t_{\rm run}$ is the runaway timescale determined numerically as the time at which $M_{\rm atm}/M_{\rm core}$ reaches 0.5:
\begin{equation}
    t_{\rm run} \sim 2.24\,{\rm Myrs} \left(\frac{20\,M_\oplus}{M_{\rm core}}\right)^{4.2} \left(\frac{Z}{0.02}\right) \left(\frac{2.37}{\mu}\right)^{8.5}.
\end{equation}
Both $Z$ and $\mu$ are evaluated at the radiative-convective boundary, assuming $Z$ is uniform throughout the planetary envelope.

The required metallicity enhancement is severe, due by the need to trigger the runaway prior to disk dissipation and reach the observed GCR (see bottom row of Figure \ref{fig:GCR_sim}). We find that the required metallicity is $Z\gtrsim 0.95$, or approximately 48 times the solar value by mass.

\subsection{Constraints from structure evolution models}

We use the same type of interior structure models as in Section \ref{ssec:gcr} in order to constrain the range of metallicities allowed by the mass and radius of WASP-107b. The planet is modeled as a well-mixed, fully-convective envelope with a radiative atmosphere and we assume a 50-50 mixture of rock and ice for metals.  We do not include a core to get a conservative (high) upper limit on the envelope metallicity \citep[see][]{Thorngren2019}. Our models are sensitive to the assumed age of the system, as the planet cools and contracts over time. We set the prior on the system age to a normal distribution corresponding to the gyrochronological age ($3.4 \pm 0.7$ Gyr; see Table \ref{tab:starParam}).

We fit for the metal mass fraction using MCMC with 6 chains in parallel.  In each chain, we burn in for 10,000 steps and then record every 10th step until we have amassed 100,000 samples.  
We infer a 3$\sigma$ upper limit of $Z\lesssim0.25$ on the envelope's metal mass as higher metallicities would result in smaller radii. Interestingly, our constraint on the metallicity ($<12.5$ times the solar value by mass at $3\sigma$) is in agreement with the metallicity constraint retrieved from the spectroscopic characterization of the 12.6 $M_\oplus$ sub-Neptune GJ 3470b \citep{Benneke2019} and suggests that the two planets may have had similar formation processes.

The 3$\sigma$ upper limit of 0.25 on the envelope metallicity is 3.8 times smaller than required by the dusty gas accretion models and effectively rules out the high-metallicity accretion scenario: the planet could not be this puffy to this day if such large amounts of metals were present in its atmosphere. We highlight the fact that the rate of dusty accretion is only weakly dependent on orbital distance, under the assumption that the dust-to-gas mass ratio is spatially constant within the disk. Our constraints on the envelope metallicity therefore rule out the dusty accretion scenario in disks with uniform dust-to-gas mass ratio. Furthermore, as mentioned above (Section \ref{ssec:gcr}), the observationally inferred GCR only provides a lower limit on the amount of gas accreted by the core, as mass-loss powered by stellar irradiation and the planet's leftover heat from formation can pare down the envelope mass fraction and is believed to sculpt the radius distribution of small planets \citep{Owen2017,Ginzburg2018, Gupta2020}. Reaching a higher GCR at formation would necessitate an even higher metallicity. 

This leaves dust-free accretion as the only plausible formation scenario for WASP-107b's massive envelope. With over 85\% of its mass contained in its H/He envelope, WASP-107b effectively provides evidence for the possibility of a $<4.6$ $M_\oplus$ core accreting extensive amounts of primordial gas.

\subsection{Dynamical consideration}\label{ssec:dyn}

Assuming that WASP-107b indeed formed as a dust-free world at $\gtrsim 1$ AU, it must have undergone inward migration, whether through the underlying disk in early, gas-rich environment or through perturbation by neighboring planets in late, gas-poor or gas-empty environment. We propose that the migration of WASP-107b happened early such that it went close to experiencing the full extent of runaway gas accretion before it was deprived from its supply of gas.

The fact that WASP-107b accreted copious amount of envelope suggests that it assembled early, beyond $\sim$1 AU. After its assembly, the planet is expected to undergo inward Type I migration \citep{Lin96,Papaloizou2007}-- this has been proposed as the formation channel of super-puffs explaining why they are frequently observed as parts of resonant chains \citep{Lee2016}. Alternatively, or additionally, WASP-107b could have arrived at its present orbit through planet-planet scattering involving WASP-107c: this scenario would have the advantage of explaining the eccentricity of planet c. Still, we note that it is uncertain whether the runaway accretion can be stopped at the right time to produce WASP-107b beyond 1 AU which may suggest that WASP-107b began its inward transport just at the right time as it reached its  envelope mass fraction of $>$85\%. 

Although our constraints on the eccentricity of WASP-107b are consistent with a circular orbit at 2$\sigma$, we find that it could have a moderate eccentricity. The timescale of tidal circularization ($\tau_{circ} = e/\Dot{e}$) for WASP-107b is $\sim$66 Myrs, much shorter than the estimated age of the system.\footnote{The circularization time is evaluated using Equation (2) of \citet{Jackson2009} using the values of $Q_p$ and $k_{2p}$ for Jupiter from \citet{Lainey2009}, with $Q'_p = Q_p/k_{2p}$  \citep{Goldreich66}, and $Q'_\star$ from Equations (1) and (2) of \citet{Penev2018}.} Any remaining eccentricity of the inner planet resulting from dynamical interactions will therefore be quickly dampened by tidal interactions with its host star. Given this short circularization timescale, a present-day eccentricity of 0.06 (median from the RV fit; see Table \ref{tab:params}) would imply recent dynamical interactions that excited the eccentricity to high values. In particular, we demonstrate using our constraints on the WASP-107b's orbit that tidal interactions alone could have brought the planet from $\gtrsim 0.10$ AU to its current orbital location over less than 1 Gyr (Figure \ref{fig:ecc_a}).

Using the occurrence of spot-crossing anomalies during planetary transits, WASP-107b was inferred to feature a high spin-orbit misalignment (40--140 degrees; \citealp{Mocnik2017,DaiWinn2017}, Rubenzahl et al., submitted). 
Such high obliquity is often considered as a sign of dynamic upheaval and could be attributed to nodal precession as was proposed for HAT-P-11b \citep{Yee2018}. Alternatively, WASP-107b is sufficiently small in mass and sufficiently far away from the star that its natal disk could have been driven to substantial spin-orbit misalignment with respect to the central star.
For a star-disk system with an outer planet that is misaligned with the disk, secular resonance between the disk's precession around this planet and the star's precession around the disk can drive a large star-disk misalignment \citep[e.g.,][]{Batygin2013,Lai2014,Zanazzi2018}. This resonance is suppressed for rapidly rotating stars in the presence of a close-in massive inner planet (i.e., star's precession around the inner planet is faster than the disk's precession around the outer planet). We confirm that for WASP-107 system, star-disk misalignment by secular resonance is possible as long as WASP-107b is beyond $\sim$0.02 AU, which it is (equation (70) of \citet{Zanazzi2018}). Another scenario that could account not only for the spin-orbit misalignment but also for WASP-107b's moderate eccentricity is that of a resonance initiated during the dispersal of an otherwise coplanar protoplanetary disk \citep{Petrovich2020}. In this case, one would expect a large mutual inclination between planets b and c.

\begin{figure}
    \centering
    \includegraphics[width=0.49\textwidth]{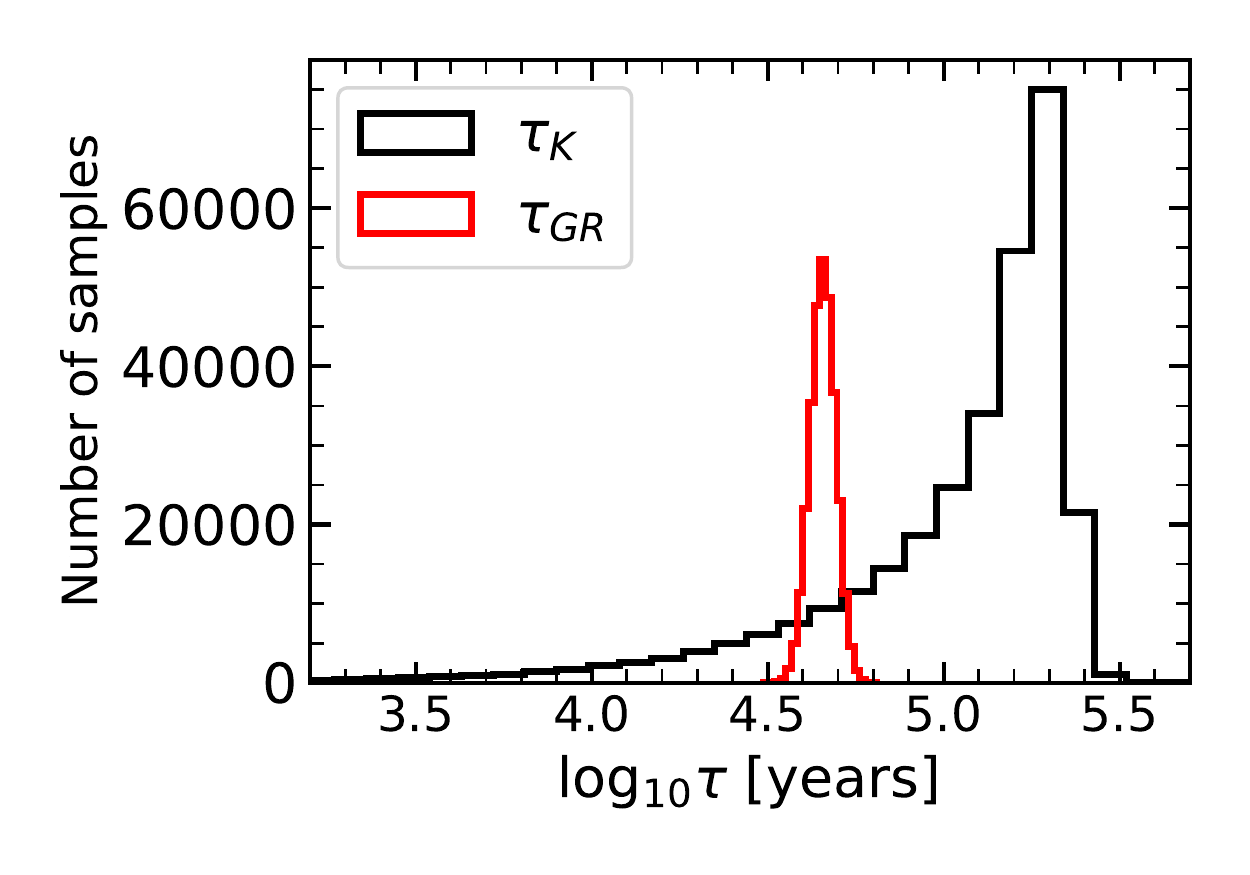}
    \caption{Distributions of the timescales of Kozai cycles ($\tau_K$) and GR apsidal precession ($\tau_{GR}$). In most cases, $\tau_K \gg \tau_{GR}$ and Kozai cycles would be efficiently suppressed.}
    \label{fig:taus}
\end{figure}
Alternatively, we examined whether Kozai-Lidov cycles induced by the presence of planet c could explain the high obliquity of WASP-107b and its moderate eccentricity. We calculate the Kozai timescale \citep{Kiseleva98} using our posterior chains and considering a random inclination for planet c. Comparing it to the period of general relativistic (GR) apsidal precession (see Figure \ref{fig:taus}), we find that the Kozai cycles are likely suppressed by GR precession, with $\tau_K \gg \tau_{GR}$.

\begin{figure}
\centering
\includegraphics[width=\linewidth]{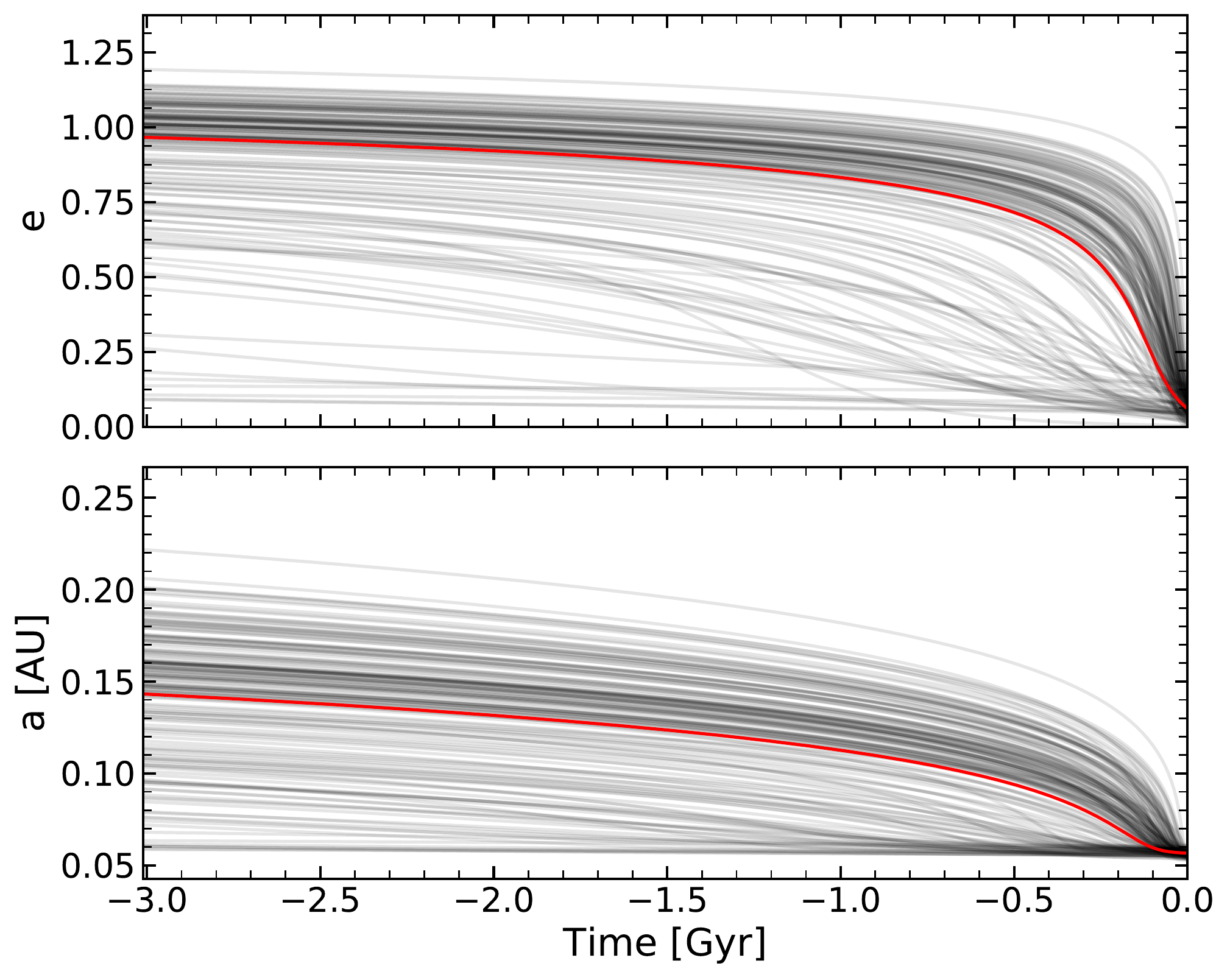}
\caption{\label{fig:ecc_a} Orbital evolution of WASP-107b (eccentricity in the top panel, semi-major axis in the bottom panel) over the past 3 Gyr, under the influence of tidal interactions from its host star alone \citep{Jackson2009}. 
The black curves are the results of 200 simulations of the orbital evolution, randomly varying relevant system parameters using their mean and 1$\sigma$ uncertainties. The red curve corresponds to a simulation using the median present-day values of each parameter.}
\end{figure}

\subsection{Tidal Inflation}
\label{ssec:caveats}

The above-mentioned scenario relies on the assumption that our two-layer models represent accurate endcases for the planetary structure. In this subsection, we consider the possibility that WASP-107b is instead a sub-Neptune with an inflated radius. 

The amount of tides raised on a planet and their dissipation can increase dramatically for planets on eccentric, inclined orbits. In particular, it was suggested that WASP-107b could be a puffed-up sub-Neptune with an envelope mass fraction as low as $\sim$10\% assuming an obliquity of $\sim$60$^\circ$, a tidal quality factor of the planet $Q'_p\sim 10^5$ and an eccentricity of $\sim$0.13 (approximately 2$\sigma$ higher than the limit we derive from our RV analysis; \citealp{Millholland2020}).

Furthermore, while we considered above the possibility that the eccentricity of WASP-107b's orbit is associated with recent dynamical interactions, it can also be explained by a large tidal quality factor for the planet. We estimate that $Q'_p \gtrsim 5\times 10^6$ is required to maintain non-zero eccentricity over the system's age against tidal circularization.
Using this $Q'_p$ and the updated eccentricity $\sim$0.06, we expect significantly less inflation by tides with a dissipation rate of $\sim$7$\times 10^{25}\,{\rm ergs\, s}^{-1}$ (approximately one order of magnitude less than the estimates of \citealp{Millholland2020}). Without running an evolutionary model, we cannot determine the expected envelope mass fraction in the presence of tidal dissipation but we estimate that it would be greater than $\sim$30--40\% based on Figure 6 of \citet{Millholland2020}, which is large enough to have initial radii of $\gtrsim 4 R_\oplus$ (i.e., a sub-Saturn). In this case, WASP-107b would have a core that weighs 18--21$M_\oplus$.  We find that in this case, its formation can be reconciled with in-situ dusty accretion with $Z\sim$0.1--0.3.

While the order of magnitude of the tidal quality factor is unconstrained observationally for this planet and higher values would lead to a lesser impact of tides on the planetary radius, tidal inflation could provide an alternative explanation for the low measured density of WASP-107b.

\section{Conclusion}\label{sec:conclusion}
With an extraordinarily low density, WASP-107b motivated both an extensive radial-velocity follow-up program and a detailed study of the physics governing the accretion of its gas envelope. 
As a first step, we use a 4-year \textit{Keck}/HIRES survey to provide a more precise measurement of WASP-107b's mass ($M_b = 30.5 \pm 1.7~M_\oplus$), 20\% (or 2.2$\sigma$) lower than the $38\pm 3~M_\oplus$ based on the CORALIE data published in the discovery paper \citep{Anderson2017}. 
The newly inferred mass indicates an extremely large envelope mass fraction of $>85$\% for solar-metallicity gas. The corresponding core mass is
$< 4.6$ $M_\oplus$ at $3\sigma$, substantially lower than  traditionally assumed to be necessary to trigger the accretion of a massive gas envelope. Higher core masses would only be possible if the planet's atmosphere had an enhanced metallicity, which we constrain to less than 12.5 times the solar value by mass at 3$\sigma$ using interior structure models. 

WASP-107b might be one of the planets that came the closest to experiencing the full extent of runaway gas accretion without becoming a gas giant. While runaway accretion was likely launched at $\gtrsim 1$ AU, the envelope growth of this super-puffy Neptune was probably stunted as a result of its migration to the inner disk. This scenario could also explain the formation of the super-puff Kepler-90g found at 0.7 AU \citep{Liang2020}.
In-situ formation, on the other hand, is ruled out for such a high gas-to-core mass ratio as it would require higher metallicities than allowed by the large radius of this planet.

We note that the orbital constraints of WASP-107b are consistent with a moderate eccentricity  (e$_b = 0.06 \pm 0.04$). If confirmed, a non-zero eccentricity could point towards recent dynamical interactions or provide an alternative explanation for its large radius via tidal inflation, provided the tidal quality factor of WASP-107b is $\lesssim 10^5$. The study of this planet thus holds the potential to inform the role that inflation plays in the observed diversity in the bulk densities of exo-Neptunes.

Our updated planetary mass, which affects the estimated surface gravity, motivates a complete re-analysis of all the published transmission spectroscopy data of WASP-107b. We are thus conducting a joint retrieval analysis of all the existing transit and eclipse spectroscopic observations, which we will present in an upcoming paper (Piaulet et al., in prep.).
Our improved understanding of the system and the bulk properties of WASP-107b will also be essential for the interpretation of JWST Guaranteed Time Observations (4 transits and 1 eclipse observation) and GO programs.

We furthermore detect a long-period more massive companion to this close-in planet on a highly eccentric orbit ($e_c = 0.28 \pm 0.07$) which may have influenced the migration and orbital obliquity of WASP-107b. 
We find that the sky-projected angular separation of planet c is of only $26_{-5}^{+8}$ mas using the posterior distributions on the period, eccentricity and argument of periastron resulting from our RV fit and assuming a random inclination of WASP-107c's orbit. This planet is thus too close to its host to be detectable via direct imaging with current instrumentation. In terms of its astrometric detectability, the reflex motion of WASP-107 as the planet travels around its orbit results in an astrometric signature of $\approx 10-30~\mu$as, smaller than the \textit{Gaia} DR2 RA/DEC uncertainty of 44 $\mu$as \citep{Gaia2018}. This suggests that WASP-107c could hardly be detected using the DR2 results but might be detectable in the future using the 5-year \textit{Gaia} time-series (Rubenzahl et al., submitted).

Looking ahead, the inferred lower mass and density of WASP-107b make it a key planet to understand and reexamine how massive a planetary core needs to be in order to accrete substantial amounts of gas.

\acknowledgements

We thank the anonymous reviewer, whose comments greatly improved the paper.
C.P. wishes to thank D. Lai and M. ali-Dib for useful discussions regarding the formation and orbital dynamics of this unique system, as well as S. Pelletier, P. Gupta, J. Chan, L.-P. Coulombe, P.-A. Roy, S. Delisle and M. Papillon for useful comments on the first manuscript of this paper. We thank D. Anderson for providing us the CORALIE RVs of WASP-107.

C.P. is supported by the Technologies for Exo-Planetary Science (TEPS) CREATE program.
C.P. and B.B. acknowledge financial support by the Fonds de Recherche Qu\'{e}b\'{e}cois—Nature et Technologie (FRQNT; Qu\'{e}bec).
B.B. further acknowledges financial support by the Natural Sciences and Engineering Research Council (NSERC) of Canada.
R. A. R. is supported by the National Science Foundation Graduate Research Fellowship Program under Grant No. DGE-1745301. Any opinions, findings, and conclusions or recommendations expressed in this material are those of the author(s) and do not necessarily reflect the views of the National Science Foundation. 
A.\,W.\,H.\ acknowledges support from the K2 Guest Observer Program and NASA's Key Strategic Mission Support program.

This research was enabled in part by support provided by Calcul Qu\'{e}bec (\url{https://www.calculquebec.ca/}), Compute Canada (\url{www.computecanada.ca}), and the Savio computational cluster resource provided by the Berkeley Research Computing program at the University of California, Berkeley (supported by the UC Berkeley Chancellor, Vice Chancellor for Research, and Chief Information Officer).

This work has made use of data from the European Space
Agency (ESA) mission Gaia (\url{https://www.cosmos.esa.int/
gaia}), processed by the Gaia Data Processing and Analysis
Consortium (DPAC, \url{https://www.cosmos.esa.int/web/gaia/
dpac/consortium}). Funding for the DPAC has been provided
by national institutions, in particular the institutions participating
in the Gaia Multilateral Agreement.

This research has made use of NASA’s Astrophysics Data System and the NASA Exoplanet Archive, which is operated by the California Institute of Technology, under contract with the National Aeronautics and Space Administration under the Exoplanet Exploration Program.

This work was based on observations obtained at the W. M. Keck Observatory, which is operated jointly by the University of California and the California Institute of Technology.

The authors wish to recognize and acknowledge the very
significant cultural role and reverence that the summit of
Maunakea has always had within the indigenous Hawaiian
community. We are most fortunate to have the opportunity to
conduct observations from this mountain.


\software{Astropy~\citep{Astropy2013},~bayev~(\url{https://github.com/exord/bayev}), corner \citep{ForemanMackey2016}, emcee~\citep{ForemanMackey2013},~isoclassify~\citep{Huber2017b}, ~Matplotlib~\citep{Hunter2007}, Numpy/Scipy~\citep{vanDerWalt2011},~RadVel~\citep{radvel}, smint~(\url{https://github.com/cpiaulet/smint}), SpecMatch-emp~\citep{Yee2017}}




\end{document}